\documentclass[%
 reprint,
showpacs,preprintnumbers,
 amsmath,amssymb,
 aps,
]{revtex4-1}

\usepackage{graphicx}
\usepackage{dcolumn}
\usepackage{bm}

\newcommand\redout{\bgroup\markoverwith{\textcolor{red}{\rule[.5ex]{2pt}{0.4pt}}}\ULon}

\usepackage[colorlinks=true,citecolor=blue]{hyperref}
\hypersetup{colorlinks=true,citecolor=blue,linkcolor=red,urlcolor=blue}

\begin{document}

\title{Charge compressibility and quantum magnetic phase transition in MoS$_2$}
\date{\today}

\author{Habib Rostami}
\affiliation{School of Physics, Institute for Research in Fundamental Sciences (IPM), Tehran 19395-5531, Iran}

\author{Reza Asgari}
\email{asgari@ipm.ir}
\affiliation{School of Physics, Institute for Research in Fundamental Sciences (IPM), Tehran 19395-5531, Iran}

\begin{abstract}
We investigate the ground-state properties of monolayer MoS$_2$ incorporating the Coulomb interaction together with a short-range intervalley interaction between charged particles between two valleys within the Hartree-Fock approximation. We consider four variables as independent parameters, namely homogeneous charge (electron or hole) density, averaged dielectric constant, spin degree of freedom and finally the Hubbard repulsion coefficient which originates mostly from $4d$ orbits of Mo atoms. We find the electronic charge compressibility within the mean-field approximation and show that non-monotonic behavior of
the compressibility as a function of carrier density which is
rather different from those of the two-dimensional electron gas. We also explore a paramagnetic-to-ferromagnetic quantum phase transition for the wide range of the electron density in the parameter space.

\end{abstract}

\maketitle
\section{introduction}\label{sect:intro}

Developments in the techniques of molecular beam epitaxy and chemical vapor deposition have allowed the fabrication of semiconductor structures in which the carriers can form a low-density fluid moving in low dimensionality~\cite{Harrison}. Many of the electron-electron interaction effects become increasingly important as carrier density or dimensionality is reduced and the homogeneous electron gas where an assembly of fermions interacting via the Coulomb interaction and moving in a uniform neutralizing background provides a primitive model for their study~\cite{March}.

Observation of elegant physical phenomena in low-dimensional systems has enticed scientists to actively explore possibilities of other two dimensional (2D) materials with outstanding characteristics. In this regard, monolayer MoS$_2$, belonging to the family of layered transition metal dichalcogenides, has been synthesized recently through mechanically cleaving bulk MoS$_2$, a layered material studied since the 1960s and which is held together by weak van der Waals interaction. Just like graphene, MoS$_2$ atoms are arranged hexagonally and it exhibits novel correlated electronic phenomena ranging from insulator to superconductor and is still flat enough to confine electrons so that charge flows quickly leading to a relatively high mobility that is promised by electronic and optical properties~\cite{wang12}.

Monolayer MoS$_2$ has recently
attracted great interest because of its potential applications in 2D
nanodevices~\cite{Mak, Radisavljevic}, owing to structural stability and the lack of dangling bonds~\cite{Banerjee}. Monolayer MoS$_2$ is a direct gap semiconductor with
an optical bandgap of $1.8$ eV~\cite{Mak}, and can be easily synthesized by using Scotch tape or
lithium-based intercalation~\cite{Mak, Radisavljevic, Banerjee, wang}. The mobility of  monolayer MoS$_2$ can be
at least $217$ cm$^2$V$^ {-1} $s$^ {-1} $ at room temperature using hafnium oxide as a high-$\kappa$ gate
dielectric, and the monolayer MoS$_2$ transistor shows room temperature current on/off
ratios of $10^8$ and ultralow standby power dissipation~\cite{Mak}. Recently, the MoS$_2$
nanoribbons have been obtained by using the electrochemical method~\cite{Li}. The
experimental achievements triggered the theoretical interest in the physical and
chemical properties of monolayer MoS$_2$ nanostructures which revealed the
origins of the observed electrical, optical, mechanical, and magnetic properties,
and guided the design of novel MoS$_2$ based devices.

Thermodynamic quantities such as the electronic compressibility, the physical observable quantity most directly related to the energy that measures the stiffness of the system against changes in the density of electrons, are a very powerful probe of exchange and correlation effects in interacting many electron systems since they are intimately linked with the equation of state~\cite{GV_book}. In an ordinary 2D electron gas, corrections to the compressibility due to the correlation effects omitted in Hartree-Fock (HF) approximations are relatively small. Ilani et al.~\cite{Ilani} performed a thermodynamic investigation of the 2D electron system measuring the compressibility. They found that the compressibility of the metallic phase largely follows Hartree-Fock theory and it is spatially homogeneous. Similar results were also reported by Dultz and Jiang~\cite{Dultz} for the thermodynamic signature of the metal-insulator transition. Moreover, compressibility measurements of 2D electron gas systems have been carried out~\cite{Eisenstein} and it is found qualitatively that Coulomb interactions affect the compressibility at sufficiently low electron density or in the strong coupling constant region. Recently, the local compressibility of graphene has also been measured~\cite{Martin} by using a scannable single-electron transistor, and theoretically the compressibility was calculated~\cite{peres}.

In recent years, because of the important
and novel physical properties found in both theoretical and
technological applications, there has been a large number of theoretical and
experimental studies on the transport properties of 2D electron systems. Although
the basic mechanism and the existence
of a quantum phase transition are still a matter of on-going debate,
experiments have amassed a wealth of data on the transport
properties of the 2D electron systems in the metallic state.
As a function of the interaction strength, which is characterized
by the ratio $r_s=(\pi n a^2_B)^{-1/2}$ in
which $a_B$ is the Bohr radius of the Coulomb energy to Fermi energy, many novels
correlated ground states have been predicted such as a paramagnetic
liquid ($r_s<26$), a ferromagnetic liquid ($26 <r_s<35$) and
Wigner crystal ($r_s>35$)~\cite{AttaccalitePRL02, reza}. A ferromagnetic~\cite{Zhang} behavior has also been reported in
MoS$_2$ and it has been related to edges or to the
existence of defects~\cite{Vojvodic}. The magnetic properties of MoS$_2$
nanoribbons indicate that the electron-electron interactions are not negligible. Furthermore, the effect of Coulomb interactions on the low-energy band structure of MoS$_2$ using an effective two-band model Hamiltonian has been recently studied~\cite{cortijo} and the study showed that a large conduction band spin splitting and a spin dependent Fermi velocity are generated due to the Coulomb interaction.

The purpose of this paper is to study the transport properties
such as band gap renormalization and charge compressibility of monolayer MoS$_2$
systems in medium and large charged densities where many-body effects are not strong. In this work, we present calculations of the zero temperature
electronic compressibility and the quantum magnetic phase transition of disorder-free monolayer MoS$_2$ based on a
two-band continuum model. We show that the compressibility of monolayer MoS$_2$ is remarkably different from the two dimensional electron gas and  from monolayer graphene. The physical behavior of the compressibility of monolayer MoS$_2$ is not a monotonic function of the charge (electron or hole) density. To investigate the magnetic phase of the ground-state in the Hartree-Fock
approximation, we use the Stoner exchange model in which it is assumed that the system is partially spin polarized. Our numerical results predict that the system with hole charge carriers can easily go to the ferromagnetic phase in contrast to the situation in which the charge carriers are electrons.

The rest of this paper is organized as follows. In the next section
we outline our theoretical approach to calculate the ground-state energy
of MoS$_2$ systems within the Hartree-Fock approximation from which the quasiparticle excitations
are obtained. The essential ingredients of our theoretical framework
are the effective inter- and intra-valley
electron-electron interactions which are discussed in Sec.\,II. Our numerical results for the bandgap renormalization and charge compressibility of both the electron and
hole-doped systems are presented
in Sec.\,III. We conclude
in Sec.\,IV with a brief summary of our main results.

\section{theory and method}
The two-band single particle Hamiltonian of monolayer MoS$_2$, neglecting the trigonal warping and the spin-orbit coupling of the conduction band, is given by~\cite{Rostami13}
\begin{eqnarray}\label{h0}
{\cal H}_0&=&\sum_{k,\tau s,\gamma \delta}{\psi^{\dagger\gamma}_{{\bf k},\tau s}{\cal H}_{\gamma\delta}({\bf k},\tau s)\psi^\delta_{{\bf k},\tau s}}\nonumber \\
{\cal H}({\bf k},\tau s)&=& \frac{\Delta}{2}\sigma_z+\lambda_{soc}\tau s
\frac{1-\sigma_z}{2}+t_0 a_0 {\bf k}\cdot{\bm \sigma}_\tau\nonumber\\
&+&\frac{\hbar^2|{\bf k}|^2}{4m_0}(\alpha+\beta\sigma_z)
\end{eqnarray}
where $a_0=0.184$nm, $\lambda_{soc}=0.08$eV, $\Delta=1.9eV$, $t_0=1.68eV$, $\alpha=0.43$, and $\beta=2.21$. Here $\gamma$ and $\delta$ refer to the conduction and valence bands. The field operators in the Hamiltonian are defined as $\psi^\dagger_{{\bf k},\tau s}=(a^\dagger_{{\bf k},\tau s},b^\dagger_{{\bf k},\tau s})$ where $a^\dagger_{{\bf k},\tau s}$ and $b^\dagger_{{\bf k},\tau s}$ are creation operators in the pseudospin space. To study the effect of electron-electron interactions, we use a model which includes both long range and short range interactions as introduced by Roldan et al~\cite{Roldan13}. We consider the interaction of quasiparticles by using
the leading diagram approximation, which is the exchange
interaction. In this sense, the interacting Hamiltonian reads~\begin{eqnarray}
{\hat V}_{long}&=&\frac{1}{2S}\sum_{{\bf q}\neq0,{\bf k}, {\bf k'},\tau s s',\gamma\delta}{v_q \psi^{\dagger\gamma}_{{\bf k}-{\bf q},\tau s} \psi^{\dagger\delta}_{{\bf k}'+{\bf q},\tau s'}\psi^{\delta}_{{\bf k}',\tau s'}\psi^{\gamma}_{{\bf k},\tau s} }\nonumber \\
&+&v_q\psi^{\dagger\gamma}_{{\bf k}-{\bf q},\tau s} \psi^{\dagger\delta}_{{\bf k}'+{\bf q},\bar\tau s'}\psi^{\delta}_{{\bf k}',\bar\tau s'}\psi^{\gamma}_{{\bf k},\tau s}\nonumber\\
{\hat V}_{short}&=&\frac{1}{2S}\sum_{{\bf k} {\bf k}'{\bf q},\tau s,\gamma\delta}{U\psi^{\dagger\gamma}_{{\bf k}-{\bf q},\tau s} \psi^{\dagger\delta}_{{\bf k}'+{\bf q},\bar\tau\bar s}\psi^{\delta}_{{\bf k}',\bar\tau\bar s}\psi^{\gamma}_{{\bf k},\tau s} }
\end{eqnarray}
where $\bar s=-s$ and $\bar\tau=-\tau$ indicating the spin and valley indices, respectively.

In order to account for screening and to avoid any divergence within Hartree-Fock theory in systems with
long-range interactions, we use a screened Hartree-Fock approach~\cite{Inkson} by generalizing an interaction potential
including Thomas-Fermi screening
\begin{eqnarray}
v_{{\bf q}}=\frac{2\pi e^2}{\epsilon_0(|{\bf q}|+\lambda q_{\rm TF})}
\end{eqnarray}
where $\epsilon_0$ is the effective dielectric constant and $q_{\rm TF}=2\pi e^2 {\cal D}(\epsilon_{\rm F})/\epsilon_0$ is Thomas Fermi screening wave vector in which ${\cal D(\epsilon_{\rm F})}=(g/2\pi)(kdk/d\varepsilon)$ is the density of states at the Fermi energy, i.e. $k=k_{\rm F}$. The parameter $\lambda$ indicates the contribution of the Thomas Fermi screening and changes between zero and unity. Notice that the Thomas Fermi wave vector is much larger than a typical Fermi wave vector due to the large band-energy effective mass that occurs in MoS$_2$. Here $g$ indicates the degeneracy of each energy level and $U=U_{4d}\times S$ where $U_{4d}$ is the Hubbard repulsion coefficient which originates mostly from $4d$ orbitals of Mo atoms~\cite{Roldan13} and $S=3\sqrt{3}/2a_0^2$ is the unit cell area.
\begin{figure}[t]
\includegraphics[width=1\linewidth]{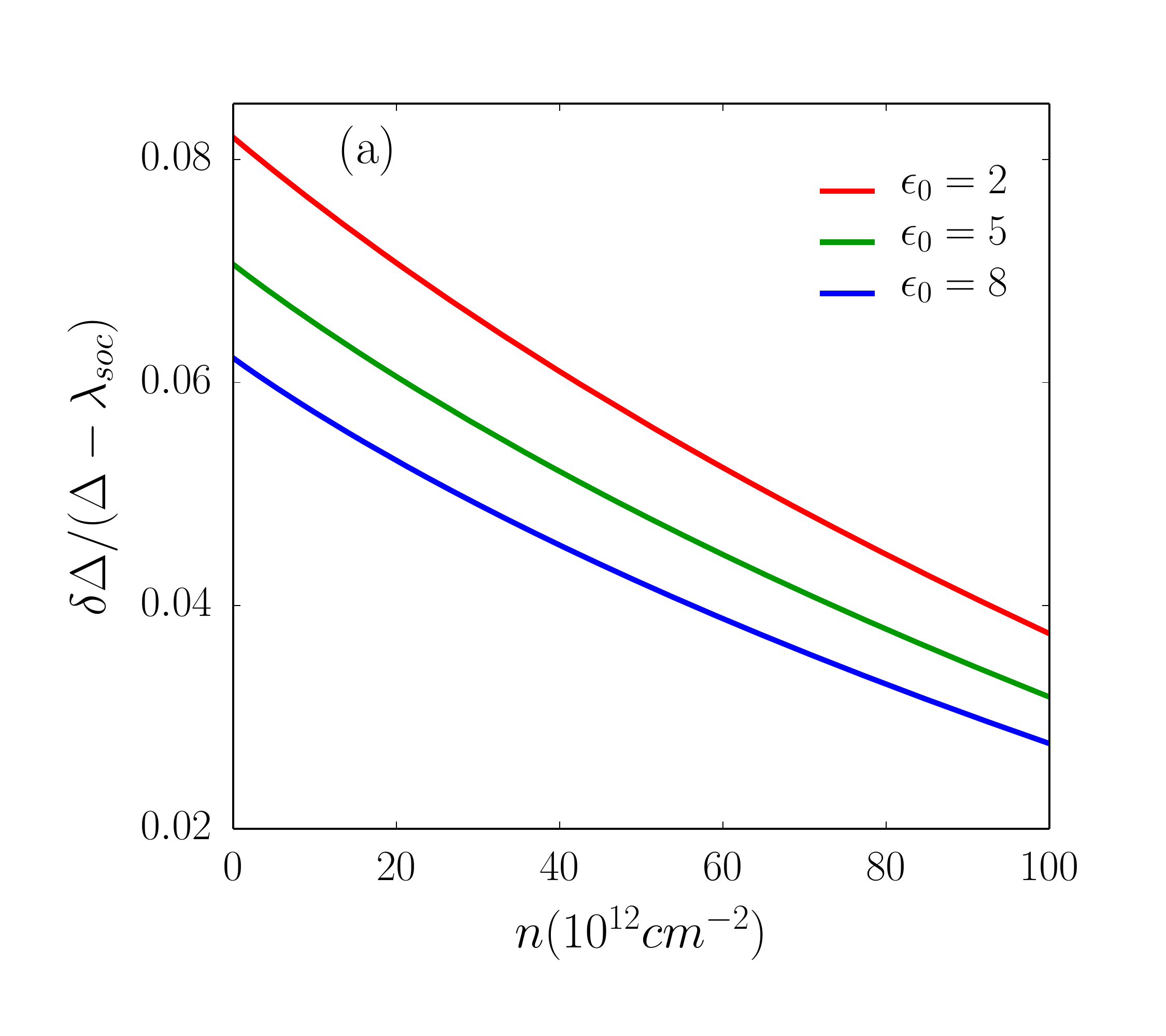}
\includegraphics[width=1\linewidth]{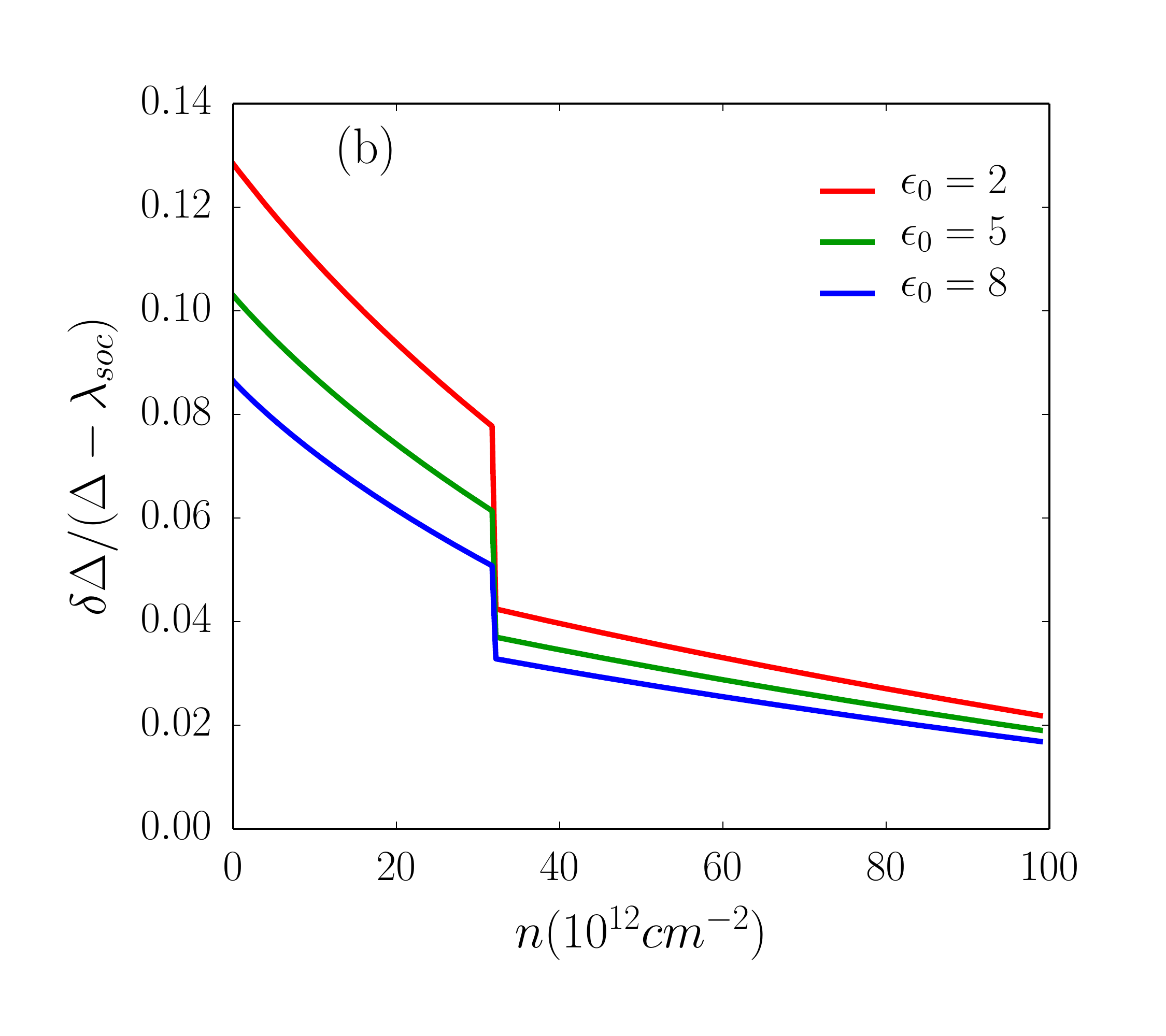}
\caption{(color online) Bandgap renormalization as a function of charge carrier density for the electron and hole doped cases for the various dielectric constants. Note that the band gap does not depend on $U_{4d}$. The band gap renormalization decreases with increasing
charge density and becomes smaller for a higher screening case. (b) in the hole doped case, The band gap renormalization shows a discontinuous function of the density associated an energy value equal to $2\lambda_{soc}$.}
\label{fig1}
\end{figure}

\begin{figure}[t]
\begin{center}
\includegraphics[width=1\linewidth]{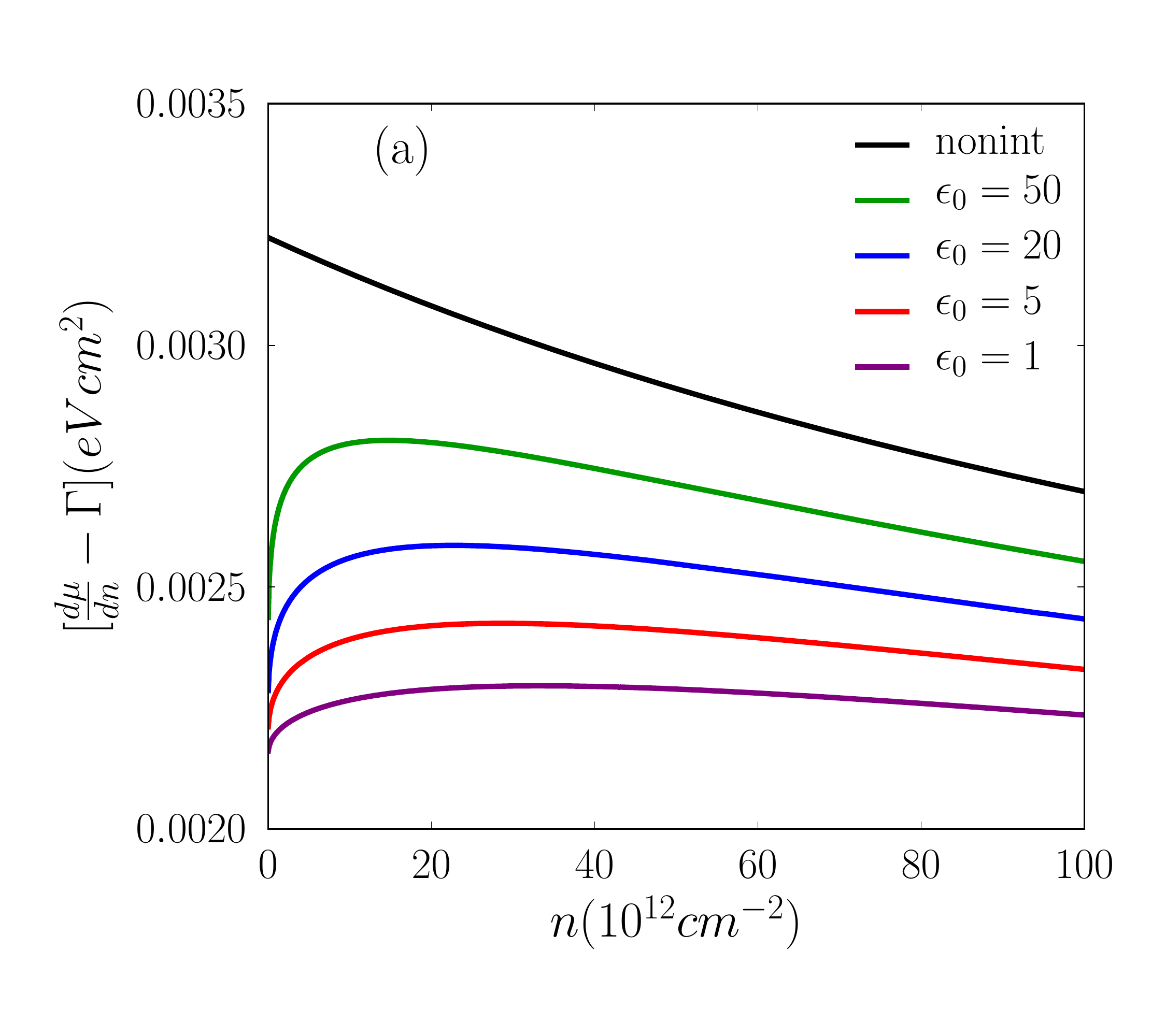}
\includegraphics[width=1\linewidth]{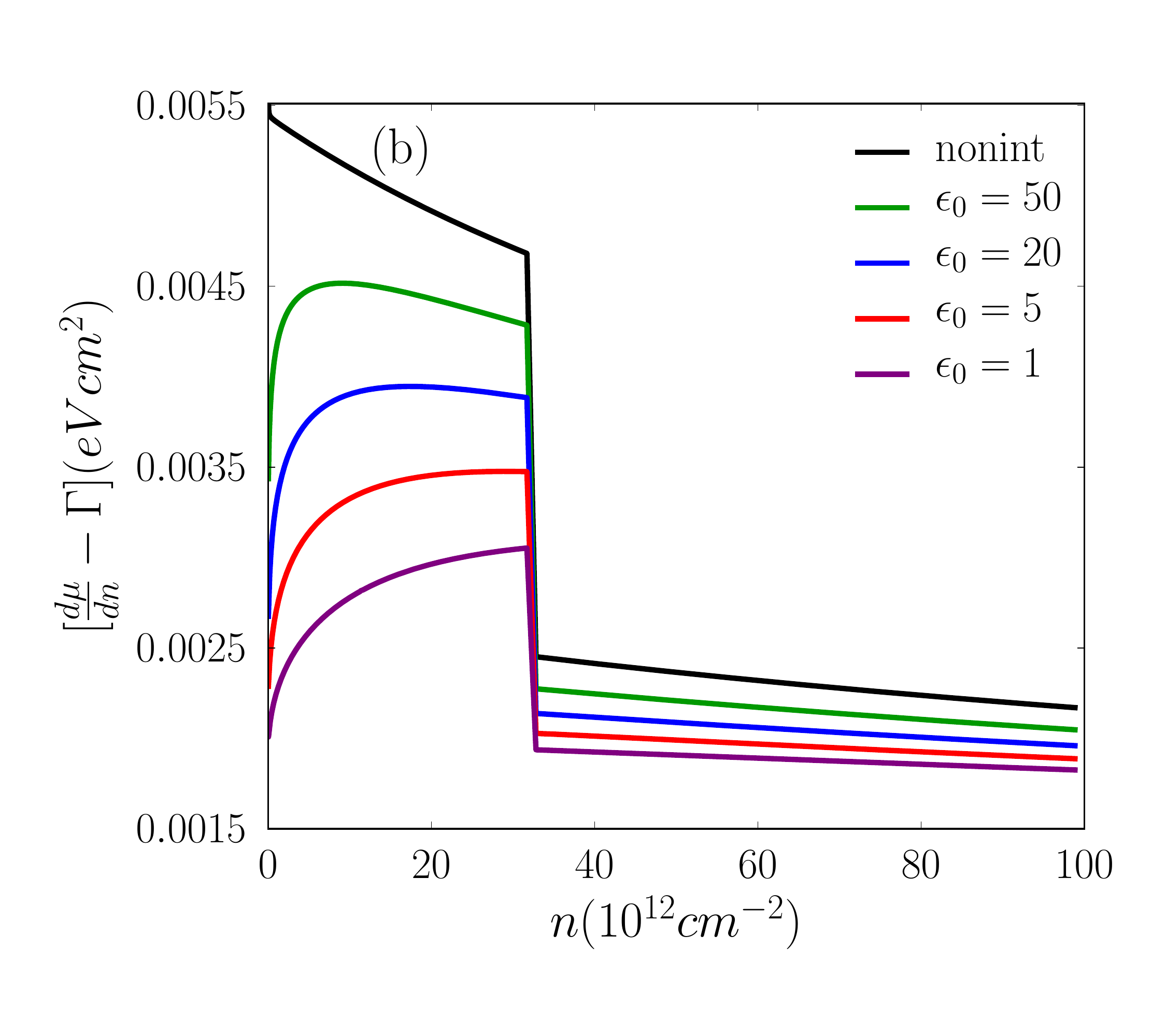}
\caption{(color online) Inverse thermodynamic density of states, $\partial \mu/\partial n$ with respect to ${\Gamma}=3\sqrt{3}a_0^2U_{4d}/80000\AA^2$ where $\mu$ is the chemical potential for (a) the electron and (b) hole doped cases as a function of the charge density for the different values of the dielectric constant. The decrease in $\partial \mu/\partial n$ with density is
a consequence of the difference between hyperbolic and
parabolic dispersion relation. We see that $\partial \mu/\partial n$ is positive and enhanced
by exchange interactions and behaves nonsymmetric with respect to the particle-hole exchange. Notice that the charge compressibility behaves non-monotonically at very low electron or hole density.}
\label{fig2}
\end{center}
\end{figure}

\begin{figure}[t]
\begin{center}
\includegraphics[width=1\linewidth]{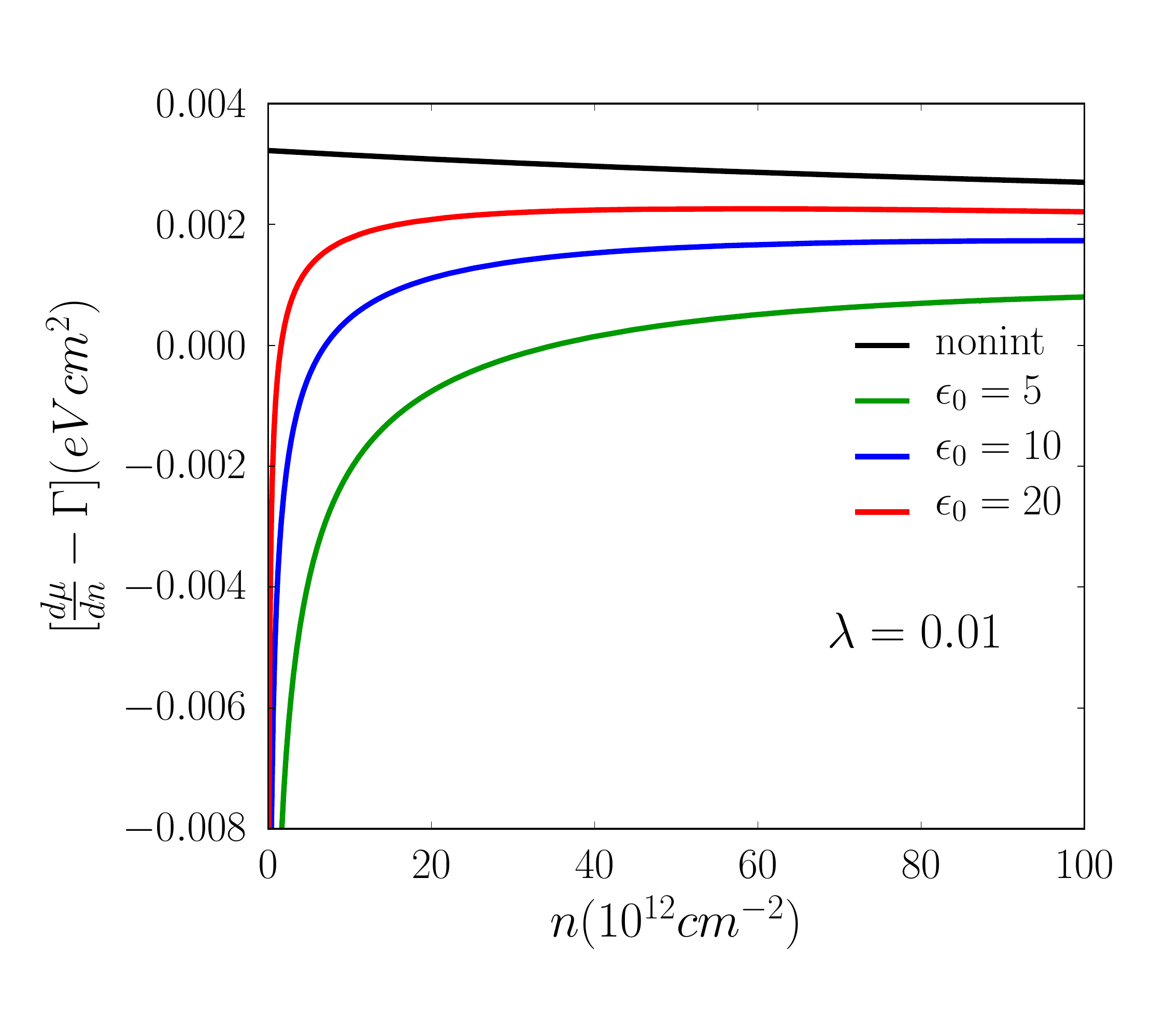}
\caption{(color online) Inverse thermodynamic density of states with respect to ${\Gamma}=3\sqrt{3}a_0^2U_{4d}/80000\AA^2$ as a function of the electron density where $\lambda=0.01$. The results are qualitatively the same as the results depicted in Fig.~\ref{fig2}.}
\label{fig3}
\end{center}
\end{figure}

\subsection{Mean field Hamiltonian}

The simplest approach to study an interacting electron gas in jellium model is the mean field
Hartree-Fock method, with the Slater determinant
wave function minimizing the ground-state energy.
In this approach, we approximate the normal-order interaction and it can be described as a neglect of the second variation of quantum fluctuations. The mean field approach provides the correct ground-state energy when the uniform electronic liquid system is pure in the absence of an external field.
In this approximation, the second term of the ${\hat V}_{long}$ vanishes. Therefore, the intravalley (long-range) and intervalley (short-range) interacting mean field Hamiltonian read ~\begin{eqnarray}
{\hat V}_{intra}&=&\frac{1}{2S}\sum_{{\bf q}\neq0,{\bf k}, {\bf k'},\tau s s',\gamma\delta}{v_q \psi^{\dagger\gamma}_{{\bf k}-{\bf q},\tau s} \psi^{\dagger\delta}_{{\bf k}'+{\bf q},\tau s'}\psi^{\delta}_{{\bf k}',\tau s'}\psi^{\gamma}_{{\bf k},\tau s} }\nonumber \\
{\hat V}_{inter}&=&\frac{1}{2S}\sum_{{\bf k} {\bf k}'{\bf q},\tau s,\gamma\delta}{U\psi^{\dagger\gamma}_{{\bf k}-{\bf q},\tau s} \psi^{\dagger\delta}_{{\bf k}'+{\bf q},\bar\tau\bar s}\psi^{\delta}_{{\bf k}',\bar\tau\bar s}\psi^{\gamma}_{{\bf k},\tau s} }
\end{eqnarray}

In unpolarized jellium, on the other hand, the common solution is a
paramagnetic state with spin symmetry. In 1962, Overhauser~\cite{Overhauser} proved that the
Hartree-Fock solution of electron gas systems is unstable with respect to spin and charge
fluctuations at any density. The global minimum energy
state within the Hartree-Fock is a spontaneously broken symmetry
state~\cite{Ceperley}.

The band eigenstates
on the positive and negative energy bands have their
pseudospins either aligned with or opposite to the direction of
the momentum~\cite{Min08,Borghi09,Rostami12}. Therefore, the mean field Hamiltonian
can be simplified as
\begin{eqnarray}
{\cal H}_{MF}&=&{\cal H}_0-\frac{1}{S}\sum_{{\bf k}{\bf k}',\tau s,\gamma\delta}{\psi^{\dagger\gamma}_{{\bf k},\tau s}v_{{\bf k}-{\bf k}'}\rho_{\gamma\delta}({\bf k}',\tau s)\psi^{\delta}_{{\bf k},\tau s}}\nonumber\\
&+&\frac{U}{S}\sum_{{\bf k} {\bf k}',\tau s,\gamma}{\rm tr[\rho({\bf k}',\bar\tau\bar s)] \psi^{\dagger\gamma}_{{\bf k},\tau s} \psi^{\gamma}_{{\bf k},\tau s}}\nonumber \\
\end{eqnarray}
where a density matrix is defined as
\begin{eqnarray}
\rho_{\gamma\delta}({\bf k},\tau s)&=&\langle\psi_0|\psi^{\dagger\delta}_{{\bf k},\tau s}\psi^{\gamma}_{{\bf k},\tau s}|\psi_0\rangle
\end{eqnarray}
Therefore, the Hartree-Fock Hamiltonian can be written as
\begin{eqnarray}
\{{\cal H}_{HF}\}_{\gamma\delta}&=&\{{\cal H}({\bf k},\tau s)\}_{\gamma\delta}-\frac{1}{S}\sum_{{\bf k}'}v_{{\bf k}-{\bf k}'}\{\rho({\bf k}',\tau s)\}_{\gamma\delta}\nonumber\\
&+&\frac{U}{S}\sum_{{\bf k}'}\rm tr[\rho({\bf k}',\bar\tau\bar s)]\delta_{\gamma\delta}
\end{eqnarray}
In order to find the density matrix, $\rho$, we calculate the eigenvalue problem of the single particle Hamiltonian i.e. ${\cal H}|\psi\rangle=E|\psi\rangle$.
As a result, we have
\begin{eqnarray}
|\psi_{\pm}\rangle&=&\frac{1}{\sqrt{(t_0a_0)^2k^2+D_\pm^2}}\begin{pmatrix}-(t_0a_0) \tau k e^{-i\tau\phi}{\rm sign}(D_\pm)\\|D\pm|\end{pmatrix} \nonumber \\
D_\pm&=&\frac{\Delta}{2}+\frac{\hbar^2 k^2}{4 m_0}(\alpha+\beta)-E_\pm\nonumber\\
E_{\pm}&=&\pm\sqrt{(\frac{\Delta-\lambda_{soc}\tau s}{2}+\frac{\hbar^2 k^2}{4 m_0}\beta)^2+(t_0a_0)^2k^2}\nonumber\\
&+&\frac{1}{2}\lambda_{soc}\tau s+\frac{\hbar^2 k^2}{4 m_0}\alpha
\end{eqnarray}
The space in which the Hamiltonian is diagonalized is based on electron ($c_k$) and hole ($v_k$) operators and $(c^\dagger_{k,\tau s},v^\dagger_{k,\tau s})=(a^\dagger_{k,\tau s},b^\dagger_{k,\tau s}){\cal U} $ where ${\cal U} $ is a unitary matrix which diagonalize the single particle Hamiltonian given by ${\cal U}=(|\psi_{+}\rangle,|\psi_{-}\rangle)$. We thus have
\begin{eqnarray}
a_{k,\tau s}&=&\frac{-(t_0a_0) \tau k e^{-i\tau\phi}}{\sqrt{(t_0a_0)^2k^2+D_\pm^2}}({\rm sign}(D_{+})c_{k,\tau s}+{\rm sign}(D_{-})v_{k,\tau s})\nonumber \\
b_{k,\tau s}&=&\frac{1}{\sqrt{(t_0a_0)^2k^2+D_\pm^2}}(|D_{+}|c_{k,\tau s}+|D_{-}|v_{k,\tau s})
\end{eqnarray}
Using the above relations together with $\langle\psi_0|c^\dagger_{k,\tau s}c_{k,\tau s}|\psi_0\rangle=n^c_{k,\tau s}$,  $\langle\psi_0|v^\dagger_{k,\tau s}v_{k,\tau s}|\psi_0\rangle=n^v_{k,\tau s}$ and $\langle\psi_0|c^\dagger_{k,\tau s}v_{k,\tau s}|\psi_0\rangle=\langle\psi_0|v^\dagger_{k,\tau s}c_{k,\tau s}|\psi_0\rangle=0$, it would be easy to find the density matrix as
\begin{eqnarray}
\rho_{aa}(k,\tau s)&=&(t_0a_0)^2k^2[\frac{n^c_{k,\tau s}}{(t_0a_0)^2k^2+D^2_{+}}+\frac{n^v_{k,\tau s}}{(t_0a_0)^2k^2+D^2_{-}}]\nonumber\\
\rho_{bb}(k,\tau s)&=&\frac{D^2_{+}n^c_{k,\tau s}}{(t_0a_0)^2k^2+D^2_{+}}+\frac{D^2_{-}n^v_{k,\tau s}}{(t_0a_0)^2k^2+D^2_{-}}\nonumber\\
\rho_{ab}(k,\tau s)&=&\rho^\ast_{ba}(k,\tau s)=-(t_0a_0)\tau k e^{-i\tau\phi}[\frac{D_{+} n^c_{k,\tau s}}{(t_0a_0)^2k^2+D^2_{+}}\nonumber\\
&+&\frac{D_{-} n^v_{k,\tau s}}{(t_0a_0)^2k^2+D^2_{-}}]\nonumber\\
\end{eqnarray}
Consequently, the mean field Hamiltonian can be written as
\begin{eqnarray}
{\cal H}_{HF}=B^{\tau s}_0({\bf k})\sigma_0+{\bf B}^{\tau s}({\bf k})\cdot {\bm \sigma}_\tau\nonumber \\
\end{eqnarray}
where
\begin{widetext}
\begin{eqnarray}\label{HF-EQ}
&&B^{\tau s}_0({\bf k})=\frac{1}{2}\lambda_{soc}\tau s+\frac{\hbar^2 k^2}{4m_0}\alpha-\frac{1}{2}\int{\frac{d^2k'}{(2\pi)^2}v_{{\bf k}-{\bf k'}}\{n^c_{k',\tau s}+n^v_{k',\tau s}\}}+U\int{\frac{d^2k'}{(2\pi)^2}\{n^c_{k'\bar\tau\bar s}+n^v_{k'\bar\tau\bar s}\}}\nonumber\\
&&B^{\tau s}_z({\bf k})=\frac{\Delta-\lambda_{soc}\tau s}{2}+\frac{\hbar^2 k^2}{4m_0}\beta-\frac{1}{2}\int{\frac{d^2k'}{(2\pi)^2}v_{{\bf k}-{\bf k'}}\{\frac{(t_0a_0)^2k'^2-D^2_{+}}{(t_0a_0)^2k'^2+D^2_{+}}n^c_{k',\tau s}+\frac{(t_0a_0)^2k'^2-D^2_{-}}{(t_0a_0)^2k'^2+D^2_{-}}}n^v_{k',\tau s}\}\nonumber\\
&& B^{\tau s}_x({\bf k})-iB^{\tau s}_y({\bf k})=(t_0a_0)\tau k e^{-i\tau \phi}+\int_0^{k_{\rm F}}\int_0^{2\pi}\frac{k'dk'd\phi'}{(2\pi)^2}v_{k-k'}{\frac{(t_0a_0)\tau k'D_{+}}{(t_0a_0)^2k'^2+D^2_{+}}e^{-i\tau\phi'}}\nonumber\\
&&~~~~~~~~~~~~~~~~~~~~~~~+\int_0^{k_c}\int_0^{2\pi}{\frac{k'dk'd\phi'}{(2\pi)^2}v_{k-k'}\frac{(t_0a_0)\tau k'D_{-}}{(t_0a_0)^2k'^2+D^2_{-}}}e^{-i\tau\phi'}
\end{eqnarray}
\end{widetext}
here $n^{c,v}_{k,\tau s}=\Theta(\varepsilon_{\rm F}-\varepsilon^{c,v}_{k,\tau s})$ is the Fermi distribution function at zero temperature. The Hamiltonian, which is main equation in the present work, consists of a
momentum dependent pseudospin effective magnetic field that acts in the direction of momentum ${\bf k}$. It must be noted that instead of performing a self-consistent procedure to find the particle distribution function we use its noninteracting expression.

\subsection{Ground-state of the electron doped system}

In order to calculate the ground-state energy within the Hartree-Fock approximation, we do need to evaluate $k_{\rm Fs} $ which is the Fermi wave vector of two spin components at each valley where they are the same, $k_{\rm F s}=k_{\rm F}(1+s\zeta)^{1/2}$, for the electron doped case however they differ from each other for a hole doped case. The Fermi wave vector given by $k_{\rm F}=\sqrt{4\pi n/g} $ where $g$ stands for the degeneracy of the band structure which is equal to 4 for the electron-doped and highly hole doped cases while in the low hole doping it is equal to 2. At zero temperature and in the electron doped case, the set of Eq.~(\ref{HF-EQ}) can be simplified as
\begin{widetext}
\begin{eqnarray}\label{h_electron}
&&~B^{\tau s}_0({\bf k})=\frac{1}{2}\lambda_{soc}\tau s+\frac{\hbar^2 k^2}{4m_0}\alpha-\frac{1}{2}\int_0^{k_{\rm F s}}\int_0^{2\pi}{\frac{k'dk'd\phi'}{(2\pi)^2}v_{{\bf k}-{\bf k'}}}+\frac{U}{4\pi}k^2_{\rm F\bar s}\nonumber\\
&&~B^{\tau s}_z({\bf k})=\frac{\Delta-\lambda_{soc}\tau s}{2}+\frac{\hbar^2 k^2}{4m_0}\beta-\frac{1}{2}\int_0^{k_{\rm F s}}\int_0^{2\pi}{\frac{k'dk'd\phi'}{(2\pi)^2}v_{{\bf k}-{\bf k'}}{\frac{(t_0a_0)^2k'^2-D^2_{+}}{(t_0a_0)^2k'^2+D^2_{+}}}-\frac{1}{2}\int_0^{k_c}\int_0^{2\pi}
{\frac{k'dk'd\phi'}{(2\pi)^2}v_{{\bf k}-{\bf k'}}\frac{(t_0a_0)^2k'^2-D^2_{-}}{(t_0a_0)^2k'^2+D^2_{-}}}}\nonumber\\
&&\tau B^{\tau s}_x({\bf k})=(t_0a_0)k\cos\phi+\int_0^{k_{\rm F s}}\int_0^{2\pi}{\frac{k'dk'd\phi'}{(2\pi)^2}v_{{\bf k}-{\bf k'}}{\frac{(t_0a_0)k'D_{+}}{(t_0a_0)^2k'^2+D^2_{+}}\cos\phi'}+\int_0^{k_c}\int_0^{2\pi}{\frac{k'dk'd\phi'}
{(2\pi)^2}v_{{\bf k}-{\bf k'}}
\frac{(t_0a_0)k'D_{-}}{(t_0a_0)^2k'^2+D^2_{-}}}\cos\phi'}\nonumber\\
&&~B^{\tau s}_y({\bf k})=(t_0a_0)k\sin\phi+\int_0^{k_{\rm F s}}\int_0^{2\pi}{\frac{k'dk'd\phi'}{(2\pi)^2}v_{{\bf k}-{\bf k'}}{\frac{(t_0a_0)k'D_{+}}{(t_0a_0)^2k'^2+D^2_{+}}\sin\phi'}+\int_0^{k_c}\int_0^{2\pi}
{\frac{k'dk'd\phi'}{(2\pi)^2}v_{{\bf k}-{\bf k'}}
\frac{(t_0a_0)k'D_{-}}{(t_0a_0)^2k'^2+D^2_{-}}}\sin\phi'}\nonumber\\
\end{eqnarray}
\end{widetext}

Here $k_c$ indicates the ultraviolet cutoff for values larger than the low-energy Hamiltonian is no longer valid and a typical value of the $k_c$ is $1/a_0$, although we set $k_c=0. 5/a_0$ to be more precise based on the comparison between the electron dispersion relation calculated by the Hamiltonian, Eq. (\ref{h0}), and those results obtained by {\it ab initio} band structure~\cite{walle12}. Notice that we ignore two infinite terms, namely $I_1=\int_0^{kc}\{kdk\}$ and $I_2=\int_0^{k_c}\int_0^{2\pi}\{v_{k-k'}kdkd\phi\}$ in the $B_0$ term which they actually originate from the integration over the whole valence bands per each spin component. It should be noted that similar simplification has been done in the case of graphene in Ref.~[\onlinecite{Borghi09}]. Moreover, the integration over $B_0$ yields
\begin{eqnarray}
\int_0^{k_{\rm F s}}{B_0^{\tau s}(k)kdk}&=&\frac{1}{4}\lambda_{soc}\tau s k^2_{\rm F s}+\frac{\hbar^2\alpha}{16m_0} k^4_{\rm F s}+
\frac{U}{8\pi}k^2_{\rm F\bar s}k^2_{\rm F s}
\nonumber\\
&-&\frac{1}{2}\int_0^{k_{\rm F s}}\int_0^{k_{\rm F s}}\int_0^{2\pi}\frac{kdkk'dk'd\phi'}{(2\pi)^2}v_{{\bf k}-{\bf k'}}\nonumber\\
\end{eqnarray}

Here, we would like to obtain analytical expressions for physical quantities when the bare Coulomb interaction is considered. To do so, we first expand the bare Coulomb interaction as
\begin{align}
v_{k-k'}=\frac{2\pi e^2}{\epsilon_0 k_{\rm F}} \sum^{\infty}_{m=-\infty} \bar V_m(x,x') e^{-i m(\phi-\phi')}
\end{align}
where $k=x k_{\rm F}$ and $k'=x' k_{\rm F}$. After straightforward calculations, the set of Eq. (\ref{h_electron}) for $\tau=+$ simplifies as
\begin{align}\label{h_electron1}
&B^{+ s}_0({\bf k})=\frac{1}{2}\lambda s+\frac{\hbar^2 k^2}{4m_0}\alpha-\frac{e^2 k_{\rm F}}{\pi\epsilon_0} I_0+\frac{U}{4\pi}k^2_{\rm F}\nonumber\\
&B^{+ s}_z({\bf k})=\frac{\Delta-\lambda s}{2}+\frac{\hbar^2 k^2}{4m_0}\beta+\frac{e^2 k_{\rm F}}{\pi\epsilon_0} I_z\nonumber\\
&[B^{+ s}_x({\bf k})-iB^{+ s}_y({\bf k})] e^{i\phi}=(t_0a_0) k +\frac{e^2 k_{\rm F}}{2\epsilon_0} I \nonumber\\
\end{align}
where
\begin{align}
&I_0=\frac{1}{x}\int_0^x x' K(\frac{x'^2}{x^2}) dx'+\int_x^1 K(\frac{x^2}{x'^2}) dx' \nonumber \\
&I_z=\frac{\pi}{2}\int_1^\Lambda x' \frac{\bar \beta k_{\rm F} x'^2+\frac{\bar\Delta}{k_{\rm F}} }
{\sqrt{(\bar \beta k_{\rm F}  x'^2+\frac{\bar\Delta}{k_{\rm F}})^2+x'^2}} \bar V_0(x,x') dx' \nonumber \\
&I=2\int_1^\Lambda x' \frac{x'}
{\sqrt{(\bar \beta  k_{\rm F} x'^2+\frac{\bar\Delta}{k_{\rm F}})^2+x'^2}} \bar V_1(x,x') dx'
\end{align}
where $K(x)$ is the elliptic integral of the first kind, $\bar\Delta=(\Delta-\lambda_{soc} s)/(t_0a_0)$ and $\bar\beta=\hbar^2\beta/(2m_0t_0a_0)$.
It is easy to find analytical expressions for $I_0$, $I_z$ and $I$ in terms of the high electron density ($x<1)$ and for the case where $\bar \beta=0$. After implementing the expressions in Eq. (\ref{h_electron1}), the renormalized value of $t_0$ is given by
\begin{align}
\frac{\tilde t_0}{t_0}-1=\gamma_0=\alpha_{ee} \ln[\frac{\Lambda+\sqrt{\Lambda^2+(\frac{\Delta}{t_0a_0k_{\rm F}})^2}}{1+\sqrt{1+(\frac{\Delta}{t_0a_0k_{\rm F}})^2}}]
\end{align}
where $\alpha_{ee} =\frac{e^2}{2\epsilon_0 t_0 a_0}$, the band gap renormalization is $\frac{\tilde\Delta}{\Delta}-1=2\gamma_0$, the spin-orbit renormalization is $\frac{\tilde\lambda_{soc}}{\lambda_{soc}}-1=2\gamma_0$, the effective mass asymmetric is renormalized as $\frac{\tilde\alpha}{\alpha}-1=\gamma_\alpha=-\frac{m_0 e^2}{2\epsilon_0\hbar^2}\frac{1-\delta_{k_{\rm F},0}}{k_{\rm F}}$ and finally the renormalized $\beta$ can be calculated as $\frac{\tilde\beta}{\beta}-1=\gamma_\beta=\alpha_{ee}\frac{m_0 a_0^2t_0^2}{\hbar^2\Delta}[\sqrt{1+(\frac{\Delta}{t_0 a_0k_{c}})^2}-\sqrt{1+(\frac{\Delta}{t_0 a_0 k_{\rm F}})^2}] $
in the Hartree-Fock approximation. The normalization of the $t$ behaves like graphene's Fermi velocity~\cite{Borghi09} by considering $\Delta=0$. Notice that the spin-dependence of $\gamma_0$, $\gamma_\alpha$ and $\gamma_\beta$ are neglected here. It is worth mentioning that the Fermi wavevector dependence of the ${\tilde\alpha}$ is similar to that result of the effective mass in 2D electron gas systems~\cite{asgari05} in the Hartree-Fock approximation.

\subsection{Ground-state of the hole doped system}
In a similar way, corresponding relations in the hole-doped case can be found as
\begin{widetext}
\begin{eqnarray}\label{h_hole}
&&~B^{\tau s}_0({\bf k})=\frac{1}{2}\lambda_{soc}\tau s+\frac{\hbar^2 k^2}{4m_0}\alpha+\frac{1}{2}\int_0^{k_{\rm F s}}\int_0^{2\pi}{\frac{k'dk'd\phi'}{(2\pi)^2}v_{{\bf k}-{\bf k'}}}-\frac{U}{4\pi}k^2_{\rm F\bar s}\nonumber\\
&&~B^{\tau s}_z({\bf k})=\frac{\Delta-\lambda_{soc}\tau s}{2}+\frac{\hbar^2 k^2}{4m_0}\beta+\frac{1}{2}\int_0^{k_{\rm F s}}\int_0^{2\pi}{\frac{k'dk'd\phi'}{(2\pi)^2}v_{{\bf k}-{\bf k'}}{\frac{(t_0a_0)^2k'^2-D^2_{-}}{(t_0a_0)^2k'^2+D^2_{-}}}-\frac{1}{2}\int_0^{k_c}\int_0^{2\pi}
{\frac{k'dk'd\phi'}{(2\pi)^2}v_{{\bf k}-{\bf k'}}\frac{(t_0a_0)^2k'^2-D^2_{-}}{(t_0a_0)^2k'^2+D^2_{-}}}}\nonumber\\
&&\tau B^{\tau s}_x({\bf k})=(t_0a_0)k\cos\phi-\int_0^{k_{\rm F s}}\int_0^{2\pi}{\frac{k'dk'd\phi'}{(2\pi)^2}v_{{\bf k}-{\bf k'}}{\frac{(t_0a_0)k'D_{-}}{(t_0a_0)^2k'^2+D^2_{-}}\cos\phi'}+\int_0^{k_c}\int_0^{2\pi}{\frac{k'dk'd\phi'}
{(2\pi)^2}v_{{\bf k}-{\bf k'}}
\frac{(t_0a_0)k'D_{-}}{(t_0a_0)^2k'^2+D^2_{-}}}\cos\phi'}\nonumber\\
&&~B^{\tau s}_y({\bf k})=(t_0a_0)k\sin\phi-\int_0^{k_{\rm F s}}\int_0^{2\pi}{\frac{k'dk'd\phi'}{(2\pi)^2}v_{{\bf k}-{\bf k'}}{\frac{(t_0a_0)k'D_{-}}{(t_0a_0)^2k'^2+D^2_{-}}\sin\phi'}+\int_0^{k_c}\int_0^{2\pi}
{\frac{k'dk'd\phi'}{(2\pi)^2}v_{{\bf k}-{\bf k'}}
\frac{(t_0a_0)k'D_{-}}{(t_0a_0)^2k'^2+D^2_{-}}}\sin\phi'}\nonumber\\
\end{eqnarray}
\end{widetext}

Notice that here there are two Fermi wave vectors $k_{\rm F1} $ and $k_{\rm F2} $ which can be calculated from
$\varepsilon^v_{k_{\rm F1},++}=\varepsilon^v_{k_{\rm F1},--}=\varepsilon_{\rm F} $ and
$\varepsilon^v_{k_{\rm F2},+-}=\varepsilon^v_{k_{\rm F2},-+}=\varepsilon_{\rm F} $, respectively, where we have used noninteracting energy dispersion. Note that $k_{\rm F2} =0$ when the Fermi energy is located in the spin splitting energy range and does not intersect with spin down (up) band around the $K$ ($K'$) point.

\subsection{Ground-state energy and quantum magnetic phase transition}

Having calculated $B_0({\bf k}) $ and $B_s({\bf k}) $, we could evaluate the energy per particle in the conduction and valence bands. The energy dispersion including the effect of electron-electron interaction is $\varepsilon^c_{k,\tau s}=B^{\tau s}_0({\bf k})+|{\bf B}^{\tau s}({\bf k}) |$ and $\varepsilon^v_{k,\tau s}=B^{\tau s}_0({\bf k})-|{\bf B}^{\tau s}({\bf k}) |$. First of all, we obtain a band gap renormalization (BGR) as a function of the electron density, which is defined as ${\rm BGR}=(\varepsilon^c-\varepsilon^v)/(\Delta-\lambda_{soc})$ where $\varepsilon^c$ and $\varepsilon^v$ indicate the conduction and valence band edges, respectively.
Notice that $B^{\tau s}_0({\bf k})$ and ${\bf B}^{\tau s}({\bf k})$ corresponding to the electron and
hole doped cases are calculated using Eqs. (12) and Eq. (14). We calculate the charge compressibility defined by $(n^2\kappa)^{-1}=\partial \mu/\partial n$ where $\mu$ is the chemical potential. The compressibility is a good quantity includes many body effects and can be measured experimentally.

In order to investigate the magnetic phase of the ground state in the Hartree-Fock approximation, we use the Stoner (or Bloch) exchange model in which it is assumed that the system is partially spin polarized. The spin polarization rate and total charge density are $\zeta=(n_\uparrow-n_\downarrow)/n$ and $n=n_\uparrow+n_\downarrow$, respectively. It should be noted that for the highly doped case, where Fermi energy intersects the spin down (up) band around the $K$ ($K'$) point as well, we have four nondegenerate bands where we should redefine $\zeta$. The total energy per particle, including the kinetic and exchange terms for an electron doped case, reads
\begin{eqnarray}
&&\varepsilon_{tot}(n,\zeta,\epsilon_0,U)=\frac{E_\uparrow+E_\downarrow}{N_\uparrow+N_\downarrow}\nonumber\\
&&E_s=\sum_{k\tau}{\varepsilon^c_{k\tau s}n^c_{k\tau s}}=\frac{S}{(2\pi)^2}\sum_{\tau}\int{\varepsilon^c_{k\tau s}n^c_{k\tau s}}d^2k\nonumber\\
&&~~~~~=\frac{S}{2\pi}\sum_{\tau}\int{\varepsilon^c_{k\tau s}n^c_{k\tau s}}kdk=\frac{S}{2\pi}\sum_{\tau}{\int_0^{k_{\rm F s}}{\varepsilon^c_{k\tau s}kdk}}\nonumber\\
&&N_s=\sum_{k\tau}{n^c_{k\tau s}}=\frac{S}{2\pi}\sum_{\tau}\int{n^c_{k\tau s}}kdk=\frac{S}{2\pi}{k^2_{\rm F s}}
\end{eqnarray}
where the total energy of the occupied state in the valence band is considered as the vacuum energy and we ignore its contribution in the energy per particle.
At zero temperature and in the electron-doped case we have
\begin{eqnarray}
\varepsilon_{tot}(n,\zeta,\epsilon_0,U)&=&\frac{\sum_{\tau s}\int_0^{k_{\rm F s}}{\varepsilon^c_{k\tau s}kdk}}{k^2_{\rm F \uparrow}+k^2_{\rm F \downarrow}}\nonumber\\
&=&\frac{1}{2k^2_{\rm F}}\sum_{\tau s}\int_0^{k_{\rm F s}}{[|{\bf B}^{\tau s}(k)|+B_0^{\tau s}(k)]kdk}\nonumber\\
\end{eqnarray}
Furthermore, for the low hole-doped case one gets
\begin{eqnarray}
\varepsilon_{tot}(n,\zeta,\epsilon_0,U)&=&-\frac{\sum_{\tau s}\int_0^{k_{\rm F s}}{\varepsilon^v_{k\tau s}kdk}}{k^2_{\rm F \uparrow}+k^2_{\rm F \downarrow}}\nonumber\\
&=&\frac{1}{2k^2_{\rm F}}\sum_{\tau s}\int_0^{k_{\rm F s}}{[|{\bf B}^{\tau s}(k)|-B_0^{\tau s}(k)]kdk}\nonumber\\
\end{eqnarray}

Finally, since the exchange interaction between itinerant electrons tends to cause a magnetic instability, the critical density~\cite{alireza} in which the paramagnetic-to-ferromagnetic Bloch phase transition~\cite{bloch} occurs can be obtained by criteria in which $\varepsilon_{tot}(n_{cr},1,\epsilon_0,U_{4d})=\varepsilon_{tot}(n_{cr},0,\epsilon_0,U_{4d})$.

Efforts to observe the ferromagnetic phase predicted by Bloch have likewise
been frustrated by the difficulty of achieving low values of the charge density. The closest thing to an
experimental observation of this transition has come so far from experiments in the 2D electron gas in a
high magnetic field. Under appropriate conditions, the magnetic field suppresses not only the kinetic energy, but also the correlation energy.
This leaves the exchange energy master of the field leads to a ferromagnetic transition. Here, we show that a magnetic transition occurs much easier in a hole-doped system than in an electron-doped case in the absence of a magnetic field.

\section{Numerical results}

We now turn to our main numerical results. The ground-state
properties of MoS$_2$ are completely determined by the
total density $n$, by the intravalley interaction $U_{4d}$ and by the media dielectric constant, $\epsilon_0$. Here, we set $\lambda=1$, otherwise we determine its value specifically.

The calculation of $\mu$ and of $\partial \mu/\partial n$ is carried out by performing
numerically the first and the second derivatives, respectively,
of the ground-state energy, which, in turn, are
known only numerically from Eqs.~(\ref{h_electron}) and (\ref{h_hole}).

\begin{figure}[t]
\includegraphics[width=1\linewidth]{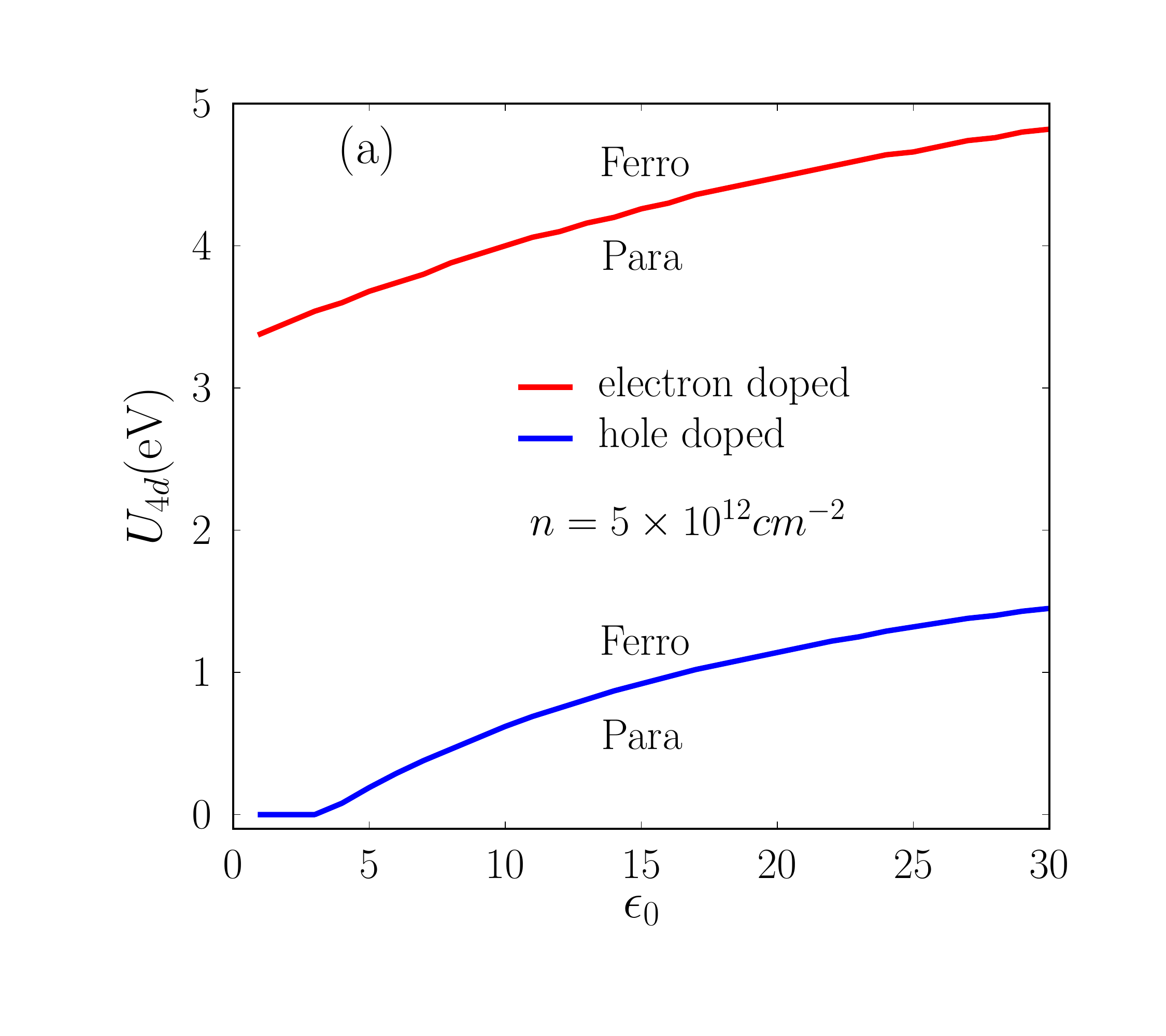}
\includegraphics[width=0.95\linewidth]{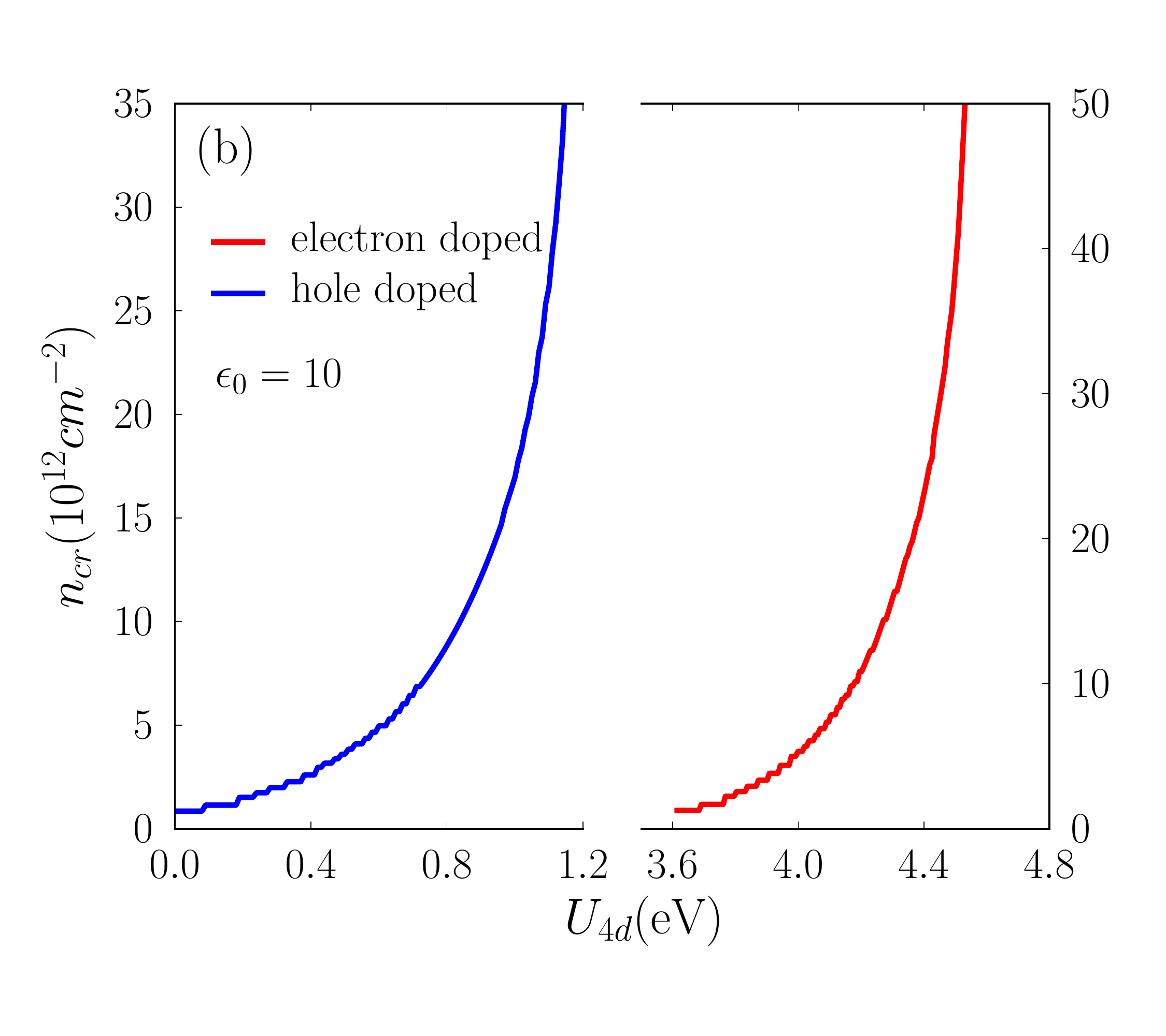}
\caption{(color online) Magnetic phase diagram in the parameters space at zero temperature. (a) $n=5\times 10^{12}cm^{-2}$ where above (below) each curve the part of the parameter space in which ground state is paramagnetic (ferromagnetic) is indicated. (b) $\epsilon_0=10$ where below (above) each curve the part of the parameter space in which the ground state is paramagnetic (ferromagnetic) is indicated.}
\label{fig4}
\end{figure}

Fig.~1 shows the BGR
for the various dielectric constants as a function of the charge density. The BGR does not depend on $U_{4d}$. The BGR decreases with increasing charge density and becomes smaller for a higher screening case. It is a smooth and monotonic function in the electron doped system shown in Fig.~1(a). However, in the hole-doped case, Fig.~1(b) we have obtained a discontinuous function of the density associated an energy value equal to the $\lambda_{soc}$ and the BGR tends to a constant weight
increasing the hole density.

In Fig.~\ref{fig2}, we report Hartree-Fock theory results in the inverse thermodynamic density of states $\partial \mu/\partial n$ with respect to $3\sqrt{3}a_0^2U_{4d}/80000\AA^2$ as a function of the charge density. The decrease in $\partial \mu/\partial n$ with density is a consequence of the difference between hyperbolic and parabolic dispersion. We see that $\partial \mu/\partial n$ is positive and enhanced by exchange over the density range covered in this plot. Since the compressibility involves only occupied states, its behavior is not symmetric with respect to particle-hole exchange. Notice that the charge compressibility behaves non-monotonically at very low electron or hole density. In Refs.~[\onlinecite{castro}] and [\onlinecite{Borghi}] a non-monotonic behavior was also found in a bilayer graphene system within the Hartree-Fock and random-phase-approximation, respectively, and the change in the sign of the inverse thermodynamic density of states predicted in very low density. This non-monotonic behavior of the compressibility as a function of carrier density is rather different from that in conventional 2D electron gas systems in which $\kappa_0/\kappa=1-r_s/2.22$ where $\kappa_0=\pi r_s^4/2$ is the compressibility of the noninteracting system and that in monolayer graphene~\cite{peres}.

We also examine our results by considering a small $\lambda=0.01$ value in which the Coulomb interaction is much larger than the screened potential particularly at the long wavelength limit. The inverse thermodynamic density of states with respect to $3\sqrt{3}a_0^2U_{4d}/80000\AA^2$ as a function of the electron density is shown in Fig.~\ref{fig3} where $\lambda=0.01$. The results are qualitatively the same as the results depicted in Fig.~\ref{fig2},
however, the value of the physical values are changed quantitatively. We find that  the change in the sign of the inverse thermodynamic density of states occurs in a larger density with decreasing $\lambda$.

To calculate the magnetic phase transition, we investigate the condition for which $\varepsilon_{tot}(n,1,\epsilon_0,U_{4d})=\varepsilon_{tot}(n,0,\epsilon_0,U_{4d})$ is satisfied by giving the $n$, $U_{4d}$ and $\epsilon_0$ parameters. It is worth to mention that in a 2D electron gas systems, the Block transition occurs at $r_s\simeq 2.01$ and there is no such transition for a massless graphene system~\cite{alireza}. Figure~\ref{fig3}(a) shows the magnetic phase diagram at a given charge density, $n=5\times 10^12$ cm$^{-2}$.
From this comparison one arrives at the conclusion that in MoS$_2$, the paramagnetic liquid has lower energy for $U_{4d} < 3$eV for the electron doped case, while the ferromagnetic liquid has lower energy for the larger value $U_{4d}$ over a wide range of the dielectric constant. Moreover, the critical value of the charge density in which the phase transition occurs is plotted as a function of the intervalley interaction for both electron and hole cases in Fig.~\ref{fig3}(b). The results suggest that the system with the hole charge carrier can easily go to the ferromagnetic phase in contract to a situation in which the charge carrier is the electron. The reason for this discrepancy is that, in the low hole doped case, the density of states is two times smaller than those of the electron doped system, owing to the spin splitting of the valence band. Consequently, the effect of screening is weaker for the holes which leads to a stronger impact of the interaction to induce a ferromagnetic phase for holes. Moreover, the spin-valley coupling of the holes results in a valley ferromagnetism together with a spin polarized magnetic phase. Therefore, the Hartree-Fock calculation predicts a spin valley polarized ground-state for the holes while that of electrons is just spin polarized in monolayer MoS$_2$.

It should be mentioned that the Bloch transition, a ferromagnetic ground-state, is not quantitatively accurate in the Hartree-Fock approximation. In order to obtain accurate ground-state energy, and a renormalized Hamiltonian for low-energy excitations
theory and to derive the expression for the
interaction function in a paramagnetic system, the knowledge of the energy functional
appropriate for an infinitesimally polarized electron liquid is needed.

\section{summary}

In conclusion, we have studied the electronic compressibility of monolayer MoS$_2$ within the Hartree-Fock approximation and have found a behavior that is remarkably different from the two dimensional electron gas and also from the
graphene monolayer. We have shown that the inverse
compressibility is not a monotonic function of the charge (electron or hole) density and it is due solely to intrinsic electronic interactions. The change in the trend of the inverse compressibility was numerically calculated and the critical value of the charge density
depends on the screening procedure that we have used in our model. We have also neglected the trigonal warping
term, which might be important at very high densities of holes. In order to investigate the magnetic phase of the ground state in the Hartree-Fock approximation, we use the Stoner exchange model in which it is assumed that the system is partially spin polarized. Our numerical results predict that the system with hole charge carriers can easily go to the ferromagnetic phase in contrast a situation in which the charge carriers are electrons.

We note that, although the Hartree-Fock method has provided
valuable information about the relative stability of the simplest phases of the electronic structure of MoS$_2$, we clearly cannot claim to have achieved a complete understanding of the magnetic phase diagram
of the system. The occurrence of transitions between states of different symmetries
indicates that the ground-state energy of the system is a nonanalytic function of parameters, namely homogeneous charge density, the averaged dielectric constant, the spin degree of freedom and finally the Hubbard repulsion coefficient.

\end{document}